\documentclass[aps,prb,twocolumn,
	amsmath,amssymb,amsfonts,floatfix,
	longbibliography,superscriptaddress]{revtex4-2}

\usepackage{graphicx} 
\usepackage{bm} 
\usepackage{amsmath}

\usepackage[hypertexnames=false,colorlinks=true,citecolor=blue,
			linkcolor=blue,urlcolor=blue]{hyperref}

\usepackage{xfrac}

\usepackage{xcolor}
\definecolor{mygreen}{rgb}{0.0, 0.6, 0.0}

\graphicspath{{fig/}{./fig/}{.}}

\begin{document}

\title{
Topological properties of spin block magnetic ladders in proximity of a superconductor: application to BaFe$_{2}$S$_{3}$
}

\author{Shivam Yadav}
\email[e-mail: ]{shivam.yadav@ifj.edu.pl}
\affiliation{Institute of Nuclear Physics, Polish Academy of Sciences, ul. E. Radzikowskiego 152, PL-31342 Krak\'{o}w, Poland}

\author{Pascal Simon}
\email[e-mail: ]{pascal.simon@universite-paris-saclay.fr}
\affiliation{Universit\'{e} Paris-Saclay, Centre National de la Recherche Scientifique, Laboratoire de Physique des Solides, 91405 Orsay, France}

\author{Andrzej Ptok}
\email[e-mail: ]{aptok@mmj.pl}
\affiliation{Institute of Nuclear Physics, Polish Academy of Sciences, ul. E. Radzikowskiego 152, PL-31342 Krak\'{o}w, Poland}

\date{\today}

\begin{abstract}
Monoatomic chains with magnetic order in proximity of a s-wave superconductor can  host  Majorana edge modes. 
In this paper, we extend this idea to more complex spin-block chains such as the BaFe$_{2}$S$_{3}$ magnetic material that has  a spin-ladder like structure. 
We investigate the topological phase diagram of such a system as function of the system parameters.
We show that the coupling between chains within the ladder leads to the topological phase with a winding number larger than the sum of two single magnetic chains.
Furthermore, strong coupling between chains leads to fractal-like substructure in the topological phase diagram.
By investigating  the real space properties of such a system and particularly its edge modes, we find that the system spectrum contains several in-gap states that we analyze in detail.
\end{abstract}

\maketitle

\section{Introduction}

The possibility to realize Majorana quasiparticles in spinless fermion chain~\cite{kitaev.01} initiated a period of intensive studies of  topological states.
Recently, several experimental setups have been proposed for the realization of the Majorana edge modes~\cite{aguado.17,lutchyn.bakkers.18,pawlak.hoffman.19}.
One of them being a magnetic chain proximitized by a s-wave superconductor~\cite{choy.edge.11,nadjperge.drozdov.13,pientka.galzman.13,braunecker.simon.13,klinovaja.stano.13,vazifeh.franz.13,braunecker.simon.15,andolina.simon.17,kaladzhyan.simon.17}.
The experimental advances have facilitated building  magnetic chains using Fe atoms~\cite{nadjperge.drozdov.14,pawlak.kisiel.16,ruby.heinrich.17,feldman.randeria.17,jeon.xie.17,kim.palaciomorales.18} or Co atoms~\cite{ruby.heinrich.17} by depositing them on a superconducting surface (like lead).
However, in such case the topological phase tends to strongly depend on the system parameters~\cite{kobialka.piekarz.20}.

The topological phase can be induced by the magnetic field in the system containing spin--orbit coupling and superconductivity~\cite{sato.takahashi.09,sato.fujimoto.09,sato.takahashi.10}.
For example, such effect were discussed in the context of  Majorana edge modes in  semiconducting nanowires with a strong spin--orbit coupling proximized by a  superconductor~\cite{bommer.hao.19, cole.sarma.15, stanescu.sarma.17}.
In such situation, the increase in the magnetic Zeeman field, leads to closing of the trivial superconducting gap, and a reopening of a ``new'' topological gap.
The effect of the external Zeeman field can be mimicked by some intrinsic  magnetic order realized within the magnetic chain, which in simplest case has ferromagnetic (FM) order.
Nevertheless, the realization of the topological phase is not limited to ferromagnetic chains, but also can be realized in the presence of antiferromagnetic (AFM) order~\cite{heimes.kotetes.14,kobialka.sedlmayr.21}.
Indeed, recent progress in  experimental techniques allow assembly of a magnetic chain by {\it atom-by-atom} engineering on top of a superconducting substrate~\cite{kim.palaciomorales.18}
The magnetic order realized within the ad-atom chain can be controlled e.g. by the relative position with respect to the surface atoms~\cite{steinbrecher.rausch.18,schneider.beck.21,schneider.beck.23,schneider.brinker.20}.
In such case, the topological phase depends strongly on the length of the chain~\cite{schneider.beck.22,schneider.brinker.20}.
In most cases, non-collinear magnetic order can support realization of the topological phase as it mimics the effect of a spin--orbit coupling and the Zeeman field~\cite{choy.edge.11,nadjperge.drozdov.13,pientka.galzman.13,braunecker.simon.13,klinovaja.stano.13,vazifeh.franz.13,braunecker.simon.15,andolina.simon.17,kaladzhyan.simon.17}.

\begin{figure}[!b]
\centering
\includegraphics[width=\linewidth]{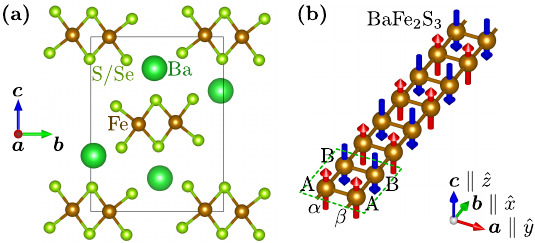}
\caption{
(a) Crystal structure of BaFe$_{2}${\it Ch}$_{3}$ ({\it Ch}=S, Se) with $Pnma$ symmetry.
(b) The iron ladder can realized canonical ($\pi$,0) AFM-like order in BaFe$_{2}$S$_{3}$.
Thus the magnetic unit cell contain two pairs of sites.
\label{fig.crys}
}
\end{figure}

In this manuscript, we  focus on the BaFe$_{2}${\it Ch}$_{3}$ ({\it Ch}=S, Se) compounds~\cite{hong.steinfink.72}.
The crystal structure consists of double chains of FeSe$_{4}$ edge connected tetrahedra intercalated with barium (Fig.~\ref{fig.crys}).
At room temperature, BaFeSe$_{3}$ exhibits multiferroic behavior~\cite{dong.liu.14,zheng.baledent.20,zheng.baledent.23} and short range magnetic order~\cite{lei.ryu.11,nambu.ohgushi.12}.
External hydrostatic pressure leads to several structural phase transition in both compounds~\cite{svitlyk.chernyshov.13,zhang.lin.18,kong.wang.19,svitlyk.garbarino.19,wy.yin.19,zheng.baledent.22}.
Moreover, the application of  pressure (between $10$ and $15$~GPa) can lead to emergence of a superconducting phase~\cite{ying.lei.17,sun.li.20}.
Finally, one of the most intriguing property is the realization of a block spin magnetic order~\cite{krztonmaziopa.pomjakushina.11,caron.neilson.11,mourigal.wu.15}, driven by electronic correlation~\cite{rincon.moreo.14,herbrych.kaushal.18,sroda.dagotto.21}.
In practice, the BaFe$_{2}$Fe$_{3}$ is a selective orbital Mott insulator~\cite{caron.neilson.12,patel.nocara.19}.
The long range magnetic order is realized below $T_{N} \sim 120$~K for BaFe$_{2}$S$_{2}$~\cite{takahashi.sugimoto.15}, and similarly below $T_{N} \sim 140$~K for BaFe$_{2}$Se$_{3}$~\cite{gao.teng.17}.
Additionally, the magnetic order can be modified by the external hydrostatic pressure~\cite{materne.bi.19,wy.yin.19,zheng.baledent.22}.

The magnetic order realized in BaFe$_{2}$S$_{3}$ and BaFe$_{2}$Se$_{3}$ correspond to the canonical $(\pi,0)$ AFM-like order [see Fig.~\ref{fig.crys}(b)].
The iron ladder exhibit FM coupling between the magnetic moments along the rungs of the ladder.
On the opposite, magnetic moments exhibit  AFM coupling along the chains of the ladder.
For BaFe$_{2}$S$_{3}$ the magnetic moments changes  orientation site-by-site [Fig.~\ref{fig.crys}(b)], while in the BaFe$_{2}$Se$_{3}$ case, they form a spin block with opposite magnetic moments.

Naturally, the presence of both a FM and AFM-like order raises the question whether topological phases could be realized in such structures when proximitized by a superconductor. 
This is the question we answer in this manuscript.
A previous theoretical investigation of the AFM chain in proximity of a superconductor shows a rich topological phase diagram~\cite{kobialka.sedlmayr.21}.

In the case of a AFM chain, additional topological regions can be realized in the phase diagram, which in some situations result in emergence of topological superconducting phases in extremely small magnetic field.
Additionally, the topological phase can be found when the Fermi level is not located around some band-edge any longer.
In our study, we specifically focus on the ladder with AFM-like order realized e.g. in BaFe$_{2}$S$_{3}$ [as presented on Fig.~\ref{fig.crys}(b)] but it can be extended to the BaFe$_{2}$Se$_{3}$ case whixh exhibits a similar phenomenology.

The paper is organized as follows.
In Sec.~\ref{sec.theo}, we first present our theoretical model describing a spin-block magnetic ladder and techniques used to study the topological phases.
Next, in Sec.~\ref{sec.results} we present and discuss our numerical results aiming at characterizing the various topological phases form a bulk or edge perspective.
Finally, we present our main conclusion of the work in Sec.~\ref{sec.sum}.


\begin{figure}[!t]
\centering
\includegraphics[width=\linewidth]{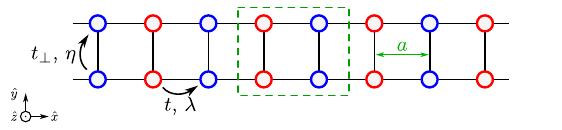}
\caption{
Schematic model of the investigated ladder mode with spin block order.
We assume the spin-conserving electron hopping along chains ($-t$) and between chains ($-t_{\perp}$).
Similarly the spin--orbit coupling is assumed along chains ($\lambda$) and between chains ($\eta$).
Lattice constant is the distance between neighbouring sites along $x$, is taken as $a = 1$.
Magnetic sublattices are present by blue and green sites, while the magnetic unit cell is marked by green dashed rectangle.
\label{fig.schem}
}
\end{figure}

\section{Model and techniques}
\label{sec.theo}

In our study we consider the magnetic ladder with AFM-like magnetic order, presented in Fig.~\ref{fig.crys}(b).
Such structure is formed by two chains (denoted $\alpha$ and $\beta$), while the magnetic unit cell contains two pairs of sites with opposite magnetic moment (with spin $\uparrow$ and $\downarrow$, called site $A$ and $B$, respectively).
Additionally, we assume that the chains are in $xy$ plane along the $\hat{x}$ direction, while magnetic moments are located along $\hat{z}$.
Thus, the hopping of electrons between sites is realized along the chains (in the $\hat{x}$ direction), and between the chains (in the $\hat{y}$ direction) (Fig.~\ref{fig.schem}).
The system can be described by the Hamiltonian:
\begin{eqnarray}
\label{eq.ham_real_space_general}
 H &=& H_{0} + H_{AFM} + H_{SC} .
\end{eqnarray}
\begin{widetext}
First term $H_0$ denotes the free electrons in the ladder:
\begin{eqnarray}
\label{eq.h0_real} H_{0} &=& -t\sum_{ \substack{i, w,\sigma}}\Big[c^{\dagger}_{i A w \sigma} c_{i B s w^{} \sigma}+c^{\dagger}_{i B w \sigma} c_{i+1 A w^{} \sigma}+h.c.\big] 
-t_{\perp}\sum_{ \substack{i, s,\sigma}}\Big[c^{\dagger}_{i s \alpha \sigma} c_{i s \beta \sigma}+h.c.\big] \\\nonumber&&
-\sum_{ \substack{i,s,w,\sigma}} \left(\mu + \sigma \, h\right)  c^{\dagger}_{i s w \sigma} c_{i s^{} w^{} \sigma} \\
\nonumber && - \imath \lambda \sum_{\substack{i  w \sigma \sigma^{\prime}}} \left( c^{\dagger}_{i A w \sigma } \sigma^{y}_{\sigma, \sigma^{\prime}} c_{i B w \sigma^{\prime}} + c^{\dagger}_{i B w \sigma } \sigma^{y}_{\sigma, \sigma^{\prime}} c_{i+1 A w \sigma^{\prime}} +h.c.\right) \\
\nonumber && - \imath \eta \sum_{\substack{i s  \sigma \sigma^{\prime}}} \left( c^{\dagger}_{i s \alpha \sigma} \sigma^{x}_{\sigma \sigma^{\prime}} c_{i s \beta \sigma^{\prime} } + h.c. \right) .
\end{eqnarray}
Here, $c_{isw\sigma}$ ($c_{isw\sigma}^{\dagger}$) denotes the annihilation (creation) operator of an electron with spin $\sigma$ in the $s=A,B$ sublattice of the $w=\alpha,\beta$ chain in the $i^{\rm th} $ unit cell. The notation $\it h.c.$ stands for hermitian conjugate.
$\mu$ and $h$ denote the chemical potential and external magnetic field, respectively.
The intra-chain and inter-chain hopping matrix elements are described by hopping integrals $t$ and $t_{\perp}$, respectively. In what follows,  $t$ will be  used as the energy unit.
Similarly, the Rashba-like spin--orbit coupling (SOC) along the $\hat{x}$ and $\hat{y}$ direction is described by the $\lambda$ and $\eta$ parameters, respectively.
The separate value of the hoppings and Rashba SOCs along and between the chains allows us to explore the large parameter space of the system and to navigate from two weakly coupled chains to a regime where the chains are strongly coupled. 
\end{widetext}

The second term describes the magnetic order:
\begin{eqnarray}
\label{eq.hafm_real} H_{AFM} &=& - m_{0} \sum_{i w \sigma} \sigma \left( c^{\dagger}_{i A w \sigma} c_{i A w \sigma} - c^{\dagger}_{i B w \sigma} c_{i B w \sigma} \right) .
\end{eqnarray}
Here, $m_{0}$ is proportional to the amplitude of the magnetic moment.
The magnetic order naturally introduces two non-equivalent sublattices $A$ and $B$ with opposite magnetic moments.
Finally, $H_{SC}$ describes the superconducting state induced within the ladder due to the proximity effect with an ordinary s-type superconductor. The pairing term reads:
\begin{eqnarray}
\label{eq.hsc_real} H_{SC} &=& \Delta \sum_{i s w} (c^{\dagger}_{i s w \uparrow} c^{\dagger}_{i s w \downarrow} + c_{i s w \downarrow} c_{i s w \uparrow}) ,
\end{eqnarray}
where $\Delta$ is taken real and corresponds to the amplitude of  the proximity induced pairing term.

\begin{widetext}
Assuming periodic boundary conditions, we can  write the Hamiltonian in momentum space using the following Fourier transform for the operators:
\begin{eqnarray}
c^{\dagger}_{i s w \sigma} = \frac{1}{\sqrt{N}} \sum_{k} \text{exp} \left( -\imath {\bm k} \cdot {\bm R}_{i s} \right) c^{\dagger}_{k s w \sigma} , \qquad \text{and} \qquad
c_{i s w \sigma} = \frac{1}{\sqrt{N}} \sum_{k} \text{exp} \left( \imath {\bm k} \cdot {\bm R}_{i s} \right) c_{k s w \sigma} ,
\end{eqnarray}
where ${\bm R}_{i s}$ denote position of $s$th sublattice site (A or B) in $i$th unit cell, i.e. ${\bm R}_{i s} = \left[ 2 i \delta_{sA} + (2 i + 1 ) \delta_{sB} \right] \hat{x}$.
Here, $c_{k s w \sigma}$ ($c^{\dagger}_{k s w \sigma}$) denotes the annihilation (creation) operator of an electron with momentum $k$ in the $s$ magnetic sublattice of the $w$ chain, while $N$ denotes the number of unit cells.
As discussed earlier the magnetic unit cell has four sites (one pair of sites per chain), which enforces the Brillouin zone (BZ) to fold on itself hence $k \in [ -\pi/2 , \pi/2)$.
In momentum space, appropriate terms of the Hamiltonian can be rewritten as:
\begin{eqnarray}
H_{0} &=& \sum_{k w \sigma} \mathcal{E}_{k} \left( c^{\dagger}_{k A w \sigma} c_{k B w \sigma} + \text{H.c.} \right) - \sum_{k s \sigma} t_{\perp} \left( c^{\dagger}_{k s \alpha \sigma} c_{k s \beta \sigma} + \text{H.c.} \right) - \sum_{k s w \sigma} (\mu + \sigma h) c^{\dagger}_{k s w \sigma} c_{k s w \sigma} \nonumber \\
&& + \sum_{k w \sigma \sigma^{\prime}} \imath \mathcal{L}_{k} \, c^{\dagger}_{k A w \sigma} \sigma^{y}_{\sigma \sigma^{\prime}} c_{k B w \sigma^{\prime}} + \text{H.c.} - \sum_{\substack{k s w \\ w^{\prime} \sigma \sigma^{\prime}}} \imath \eta \, c^{\dagger}_{k s w \sigma} \sigma^{x}_{\sigma \sigma^{\prime}} c_{k s w^{\prime} \sigma^{\prime}} + \text{H.c.} , \\
H_{AFM} &=& - m_{0} \sum_{k w \sigma} \sigma (c^{\dagger}_{k A w \sigma} c_{k A w \sigma} - c^{\dagger}_{k B w \sigma} c_{k B w \sigma}) , \\
H_{SC} &=& \Delta \sum_{k s w} (c^{\dagger}_{k s w \uparrow} c^{\dagger}_{-k s w \downarrow} + c_{-k s w \downarrow} c_{k s w \uparrow}) .
\end{eqnarray}
\end{widetext}
Here, $\mathcal{E}_{k} = -2t \cos k$, and $\mathcal{L}_{k} = -2 \imath \sin k$.
As we can see, the presented model can also be interpreted as two-orbital ($\alpha$ and $\beta$ wires) system with hybridization $t_{\perp}$ and $\eta$ between orbitals at the same ($A$ and $B$) sites.

We wish to write the Hamiltonian in the basis of tensor product of four Pauli matrices. We therefore introduce $\tau$ which acts on particle hole subspace, $\rho$ for the sublattice subspace, $\nu$ representing the chain subspace, and $\sigma$ which acts in the spin subspace. 
Using such a representation the Hamiltonian reads:
\begin{eqnarray}
H = \sum_k \psi^{\dagger}_k \mathbb{H}_k \psi_k ,
\end{eqnarray}
where $\mathbb{H}_k$ is the Hamiltonian in the matrix form for momentum $k$, while $\psi^{\dagger}_k$ is the Nambu vector for such a representation is given by:
\begin{widetext}
\begin{eqnarray}
&& \psi^{\dagger}_k = \\
\nonumber && \left(
c^{\dagger}_{k A \alpha \uparrow} \; c^{\dagger}_{k A \alpha \downarrow} \; c^{\dagger}_{k A \beta \uparrow} \; c^{\dagger}_{k A \beta \downarrow} \; c^{\dagger}_{k B \alpha \uparrow} \; c^{\dagger}_{k B \alpha \downarrow} c^{\dagger}_{k B \beta \uparrow} \; c^{\dagger}_{k B \beta \downarrow} \; c_{-k A \alpha \uparrow} \; c_{-k A \alpha \downarrow} \; c_{-k A \beta \uparrow} \; c_{-k A \beta \downarrow} \; c_{-k B \alpha \uparrow} \; c_{-k B \alpha \downarrow} \; c_{-k B \beta \uparrow} \; c_{-k B \beta \downarrow}
\right)
\end{eqnarray}
The Hamiltonian in matrix form can be represented as:
\begin{eqnarray}
\label{eq.ham_mom_matrix} && \mathbb{H}_{k} = \\
\nonumber && \mathcal{E}_{k} \tau^{3} \rho^{1} \nu^{0} \sigma^{0} - t_{\perp} \tau^{3} \rho^{0} \nu^{1} \sigma^{0} - \mu \tau^{3} \rho^{0} \nu^{0} \sigma^{0} - h \tau^{3} \rho^{0} \nu^{0} \sigma^{3} + \imath \mathcal{L}_{k} \tau^{3} \rho^{1} \nu^{0} \sigma^{2} + \eta \tau^{3} \rho^{0} \nu^{2} \sigma^{1} - \Delta \tau^{2} \rho^{0} \nu^{0} \sigma^{2} - m_{0} \tau^{3} \rho^{3} \nu^{0} \sigma^{3} \,,
\end{eqnarray}
the superscript labels which ranges from 0 to 3 for all subspaces represent the identity, $x$, $y$, and $z$ Pauli matrices respectively in their respective subspaces.
\end{widetext}

Similar to the previous study~\cite{tewari.sau.12} , the topological properties can be studied using the $\mathbb{Z}$ topological invariant.
In the case of chain-like system the winding number is the corresponding topological invariant, and can be introduced by transforming the Hamiltonian into a block form with a unitary transformation, $\mathcal{U} = \text{exp}(i \frac{\pi}{4} \tau^{3}) \rho^0 \nu^0 \sigma^0$. 
This morphs the Hamiltonian into the following block off diagonal form~\cite{tewari.sau.12}:
\begin{eqnarray}
\label{eq.ham_off_diag} \mathbb{H}_{k} &=& \begin{pmatrix}
0 & \mathcal{A}(k) \\
\mathcal{A}^{\dagger}(-k) & 0
\end{pmatrix} .
\end{eqnarray} 
The winding number now reads as
\begin{eqnarray}
\label{eq.winding} \mathcal{W} = \frac{1}{2 \pi \imath} \int_{k = -\frac{\pi}{2}}^{k = \frac{\pi}{2}} \, \frac{dz_{k}}{ z_k }
\end{eqnarray}
here $z_k = \text{Det}(\mathcal{A})$, and we note that $\mathcal{A}(k) =  \mathcal{E}_{k} \rho^{1} \nu^{0} \sigma^{0} - t_{\perp} \rho^{0} \nu^{1} \sigma^{0} - \mu \rho^{0} \nu^{0} \sigma^{0} - h \rho^{0} \nu^{0} \sigma^{3} + i \mathcal{L}_{k} \rho^{1} \nu^{0} \sigma^{2} + \eta \rho^{0} \nu^{2} \sigma^{1} - m_{0} \rho^{3} \nu^{0} \sigma^{3} + i \Delta \rho^{0} \nu^{0} \sigma^{2}$ is the $8 \times 8$ matrix.

\section{Results and discussion}
\label{sec.results}

\begin{figure}[!t]
\centering
\includegraphics[width=\linewidth]{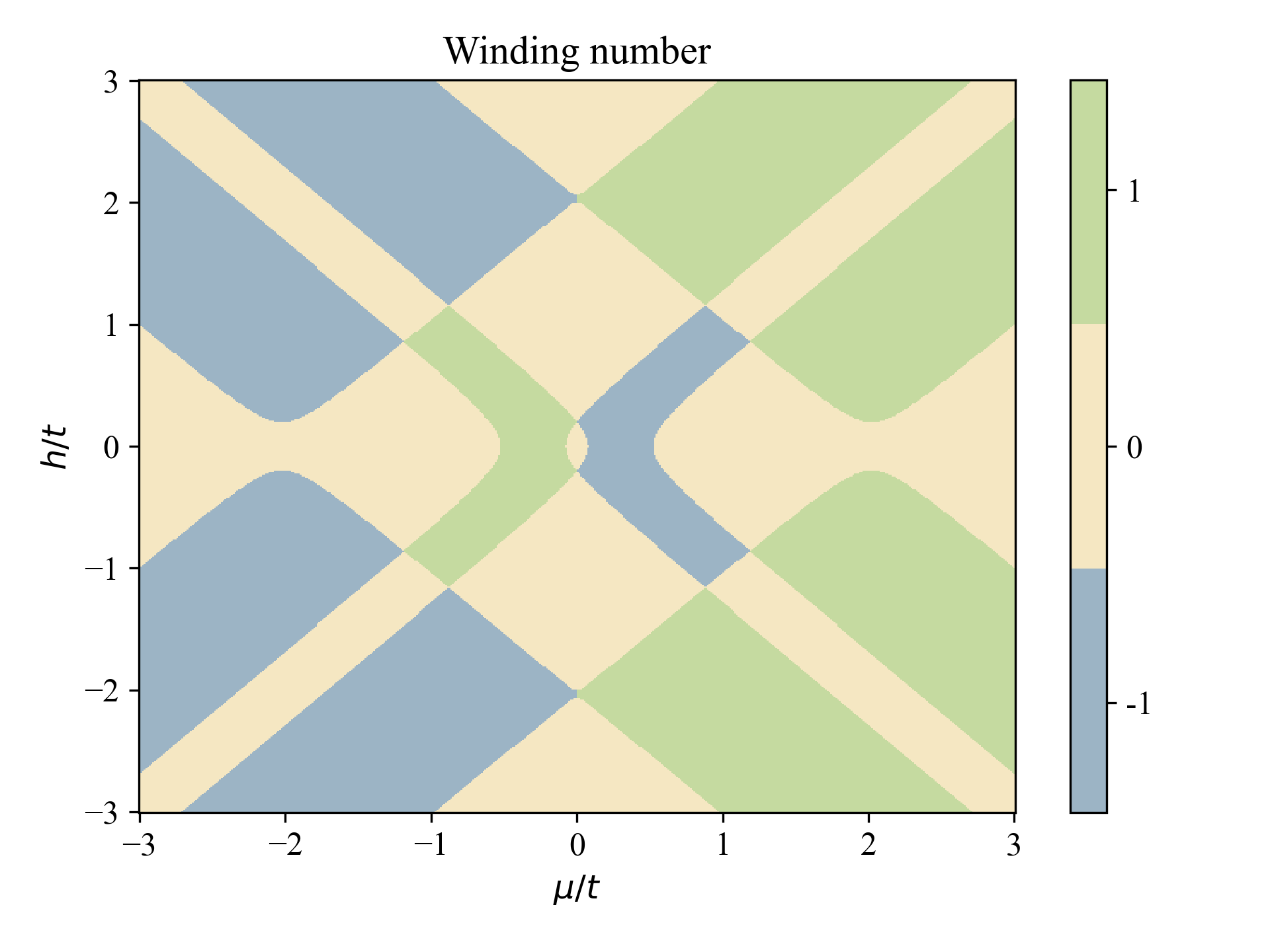}
\caption{The topological phase diagram of the one dimensional AFM chain ($t_{\perp} = \eta = 0$).
Results obtained for $m_0 /t = 0.3$, $\Delta / t = 0.2$, and $\lambda / t = 0.15$, correspond to the study presented in Ref.~\cite{kobialka.sedlmayr.21}.
\label{fig.1dafm_chain}
}
\end{figure}

\begin{figure}[!b]
\centering
\includegraphics[width=\linewidth]{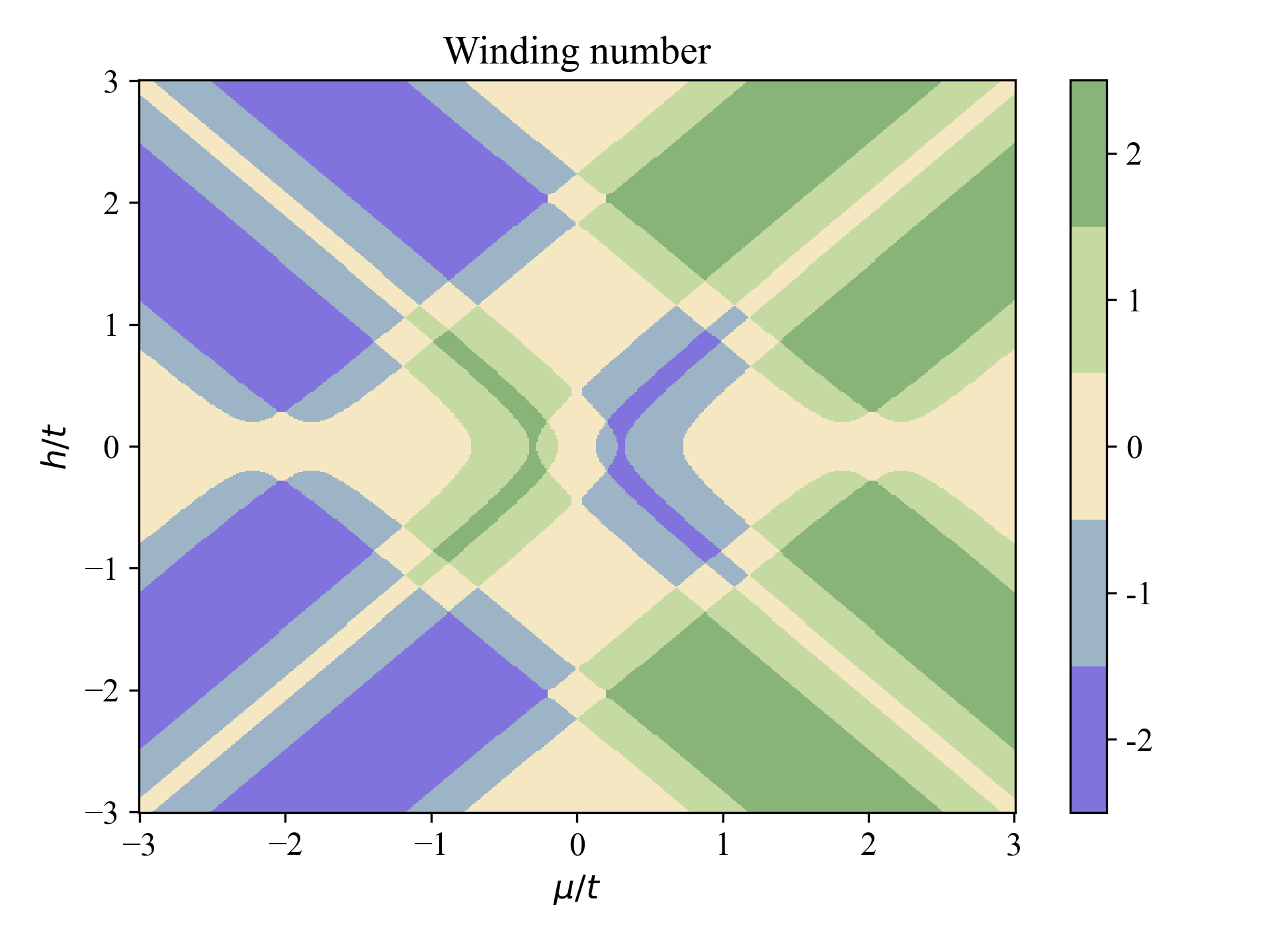}
\caption{
The topological phase diagram of the ladder with AFM-like order.
Results for the same parameters as Fig.~\ref{fig.1dafm_chain}, while coupling between chains within ladder is set as $t_{\perp} / t = 0.2$ and $\eta / \lambda = 0.2$.
\label{fig.hmu_ladder}
}
\end{figure}

Let us briefly review the results obtained for a ideal one dimensional (1D) chain with AFM order.
Here we focus on the study presented in Ref.~\cite{kobialka.sedlmayr.21}.
This system corresponds to the one presented in Sec.~\ref{sec.theo} with $t_{\perp} = \eta = 0$ (accompanied by removal of $\nu$ basis from the tensor product).
The topological phase diagram can be determined by the winding number $\mathcal{W}$.
In the absence of magnetic order, the edge of topological region can be well described by the $h_{c} ( \mu ) = \sqrt{ ( - 2 t - \mu )^{2} + \Delta^{2} }$~\cite{sato.fujimoto.09,sato.takahashi.09,sato.takahashi.10}.
For fixed $\mu$, the topological phase is realized for $\mid h \mid > h_{c} ( \mu )$, which corresponds to the topologically non-trivial regions in the form of parabolas in $h$--$\mu$ phase diagram.
However, the introduction of AFM magnetic order leads to the emergence of new region in the phase diagram (along diagonal directions $h = \pm \mu$, as shown on Fig.~\ref{fig.1dafm_chain}).
Due to  negative ``interference'' with the old topological region, the new trivial part occur inside it. 
Simultaneously, new parts of topological region can be found outside of the old topological region.
In practice, the size of such regions is controlled by the amplitude of magnetic moments. 
When the magnetic moments are large enough, the topological regions can be realized even at half-filling ($\mu = 0$) and in absence of external magnetic field ($h = 0$).

\begin{figure}[!t]
\centering
\includegraphics[width=\linewidth]{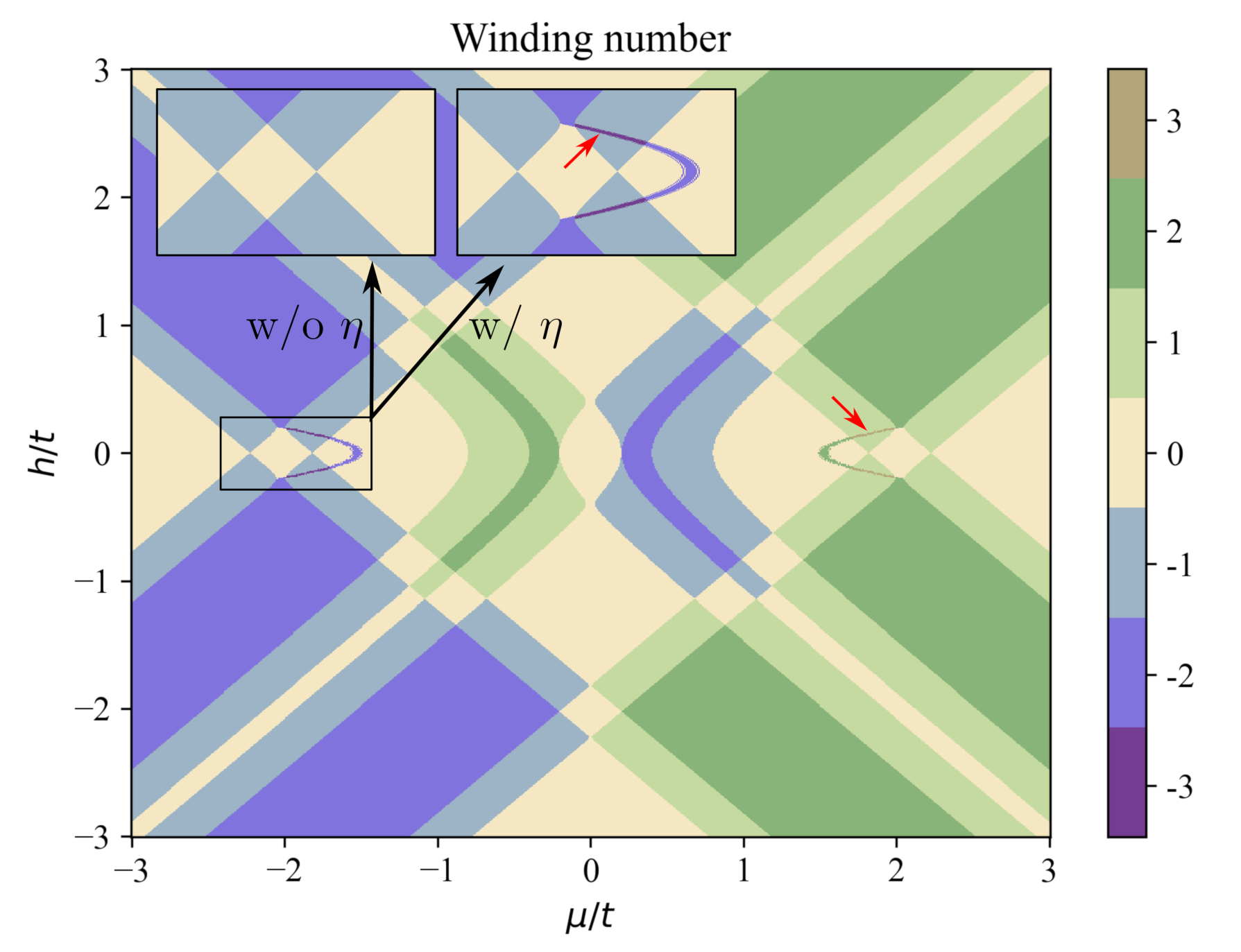}
\caption{
The same as Fig.~\ref{fig.hmu_ladder} for $\Delta / t = 0.0015$.
Inset show zoom of marked area in the absence and presence of the inter-chain spin--orbit coupling $\eta$ (left and right inset, respectively).
\label{fig.hmu_ladder2}
}
\end{figure}

\begin{figure}[!b]
\centering
\includegraphics[width=\linewidth]{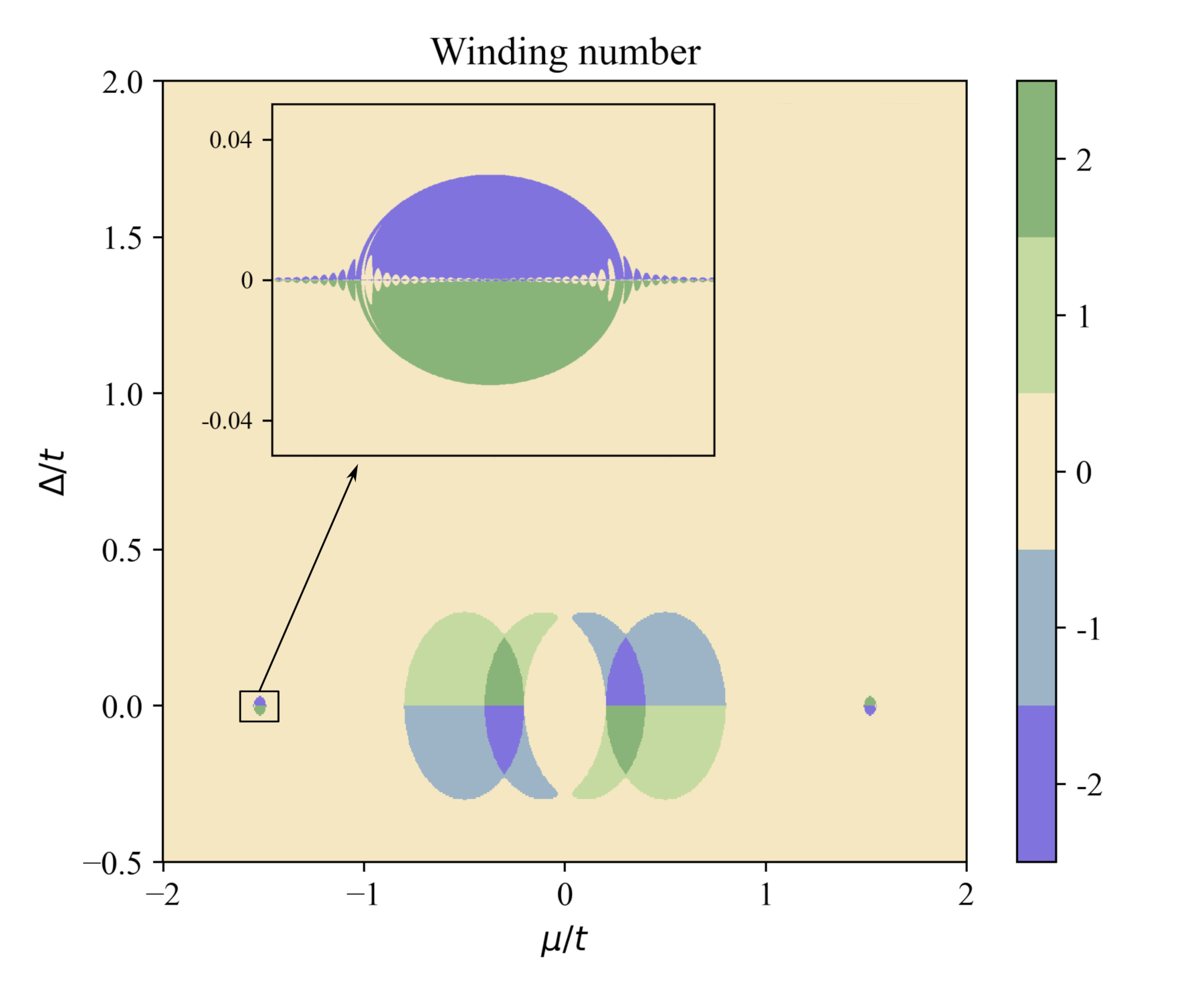}
\caption{Crossection of the topological phase diagram presented on Fig.~\ref{fig.hmu_ladder} for $h = 0$.
\label{fig.hmu_ladder3}}
\end{figure}

\begin{figure*}
\centering
\includegraphics[width=\linewidth]{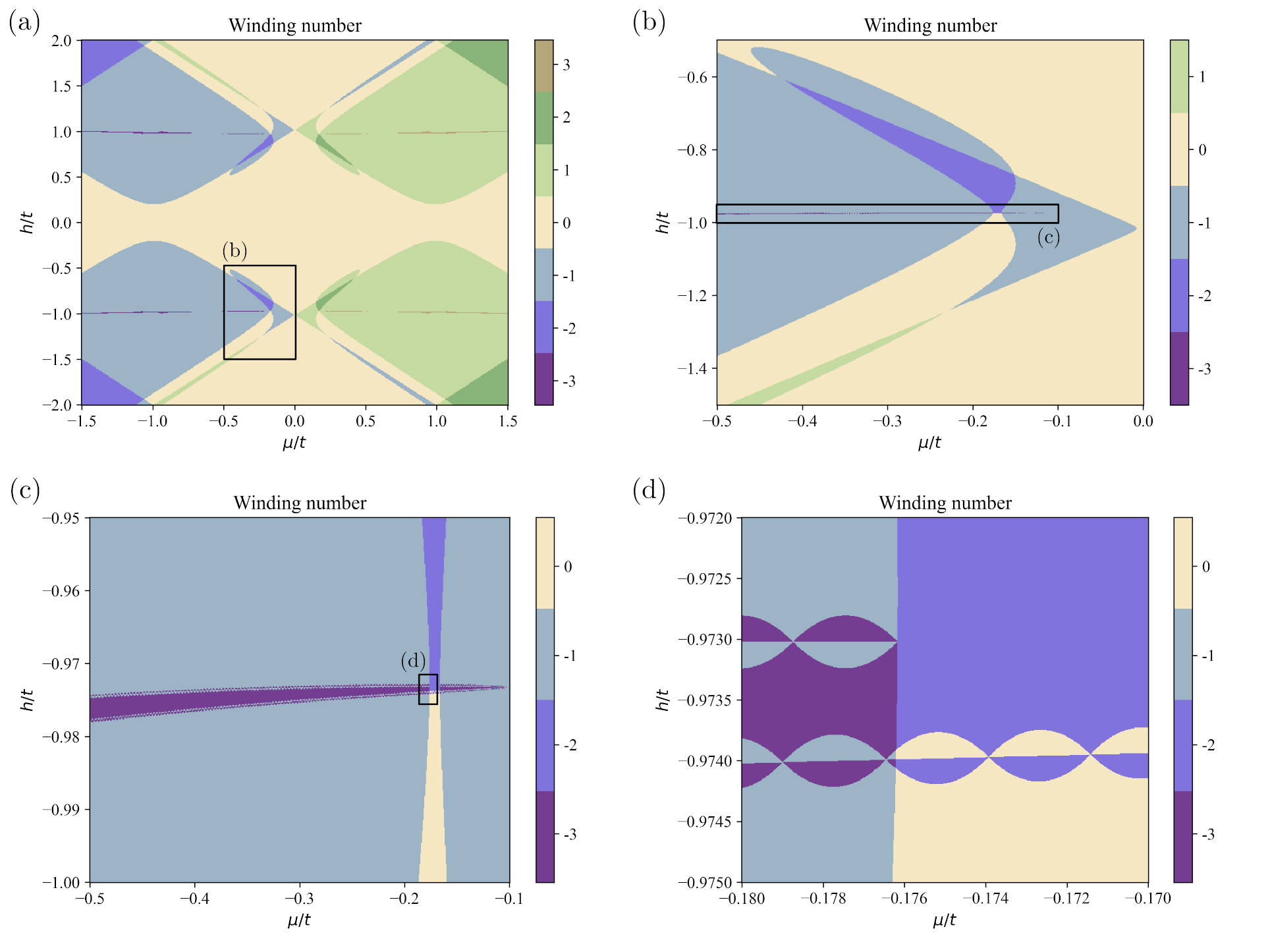}
\caption{The topological phase diagram of the ladder with AFM-like order.
Next panels show zoom on marked area (as labeled).
Results obtained for $t_{\perp} / t = 1$, $\Delta / t = 0.2$, $m_0 / t = 0.1$, and $\lambda = \eta = -0.15 t$.
\label{fig.hmu_ladder3}
}
\end{figure*}

\begin{figure}[!t]
\centering
\includegraphics[width=\linewidth]{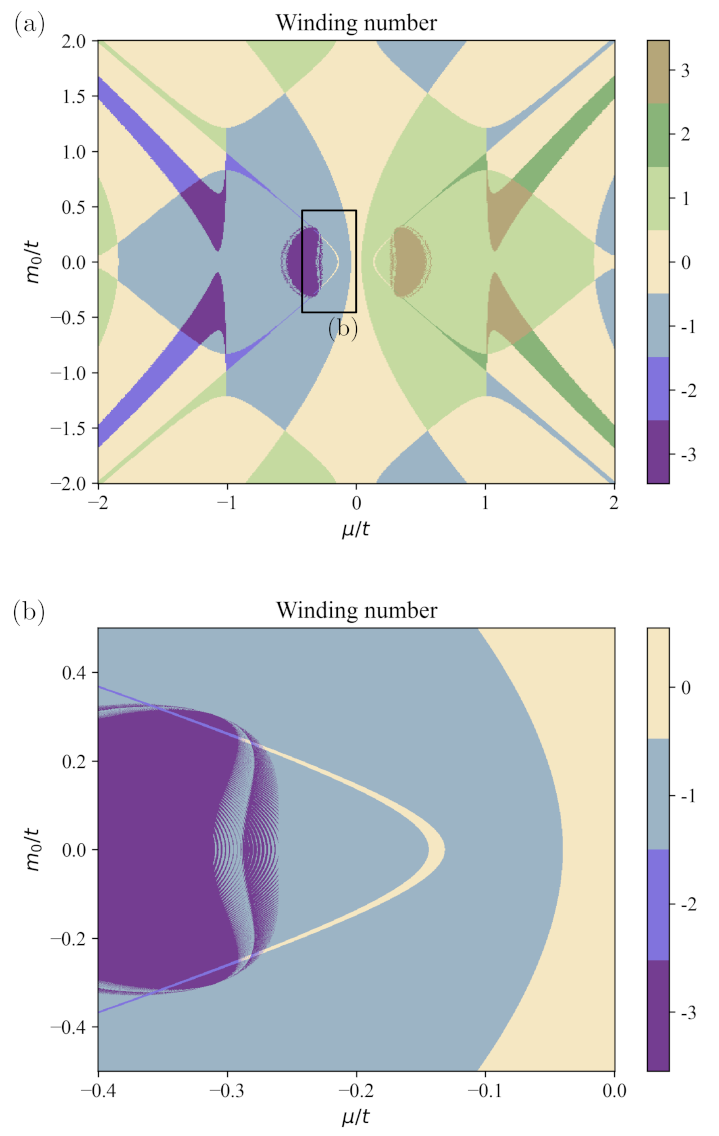}
\caption{The topological phase diagram of the ladder with AFM-like order.
Panel (b) show show zoom on marked area at panel (a).
Results obtained for $t_{\perp} / t = 1$, $\Delta / t = 0.2$, $h / t = -0.975$, and $\lambda = \eta = -0.15 t$.
\label{fig.hmu_ladder4}
}
\end{figure}

\subsection{Topological phase diagrams for ladder}

The topological phase diagram for ladder in the presence of relatively weak couplings (associated with the parameters $t_{\perp}$ and $\eta$) between the chains within the ladder is presented in Fig.~\ref{fig.hmu_ladder}.
Results are obtained using the same parameters for the  chains as the ones used  in Fig.~\ref{fig.1dafm_chain}, while the interchain couplings is set to $t_{\perp} = 0.2 t$ and $\eta = 0.2 \lambda$.
As we can see, the phase diagram exhibits topological regions ``splitted'' with respect to the ones observed for the 1D AFM chain [cf.~Fig.~\ref{fig.1dafm_chain} and Fig.~\ref{fig.hmu_ladder}].
Such splitting is associated with a shift of the topological phase boundary within the chemical potential channel.
Additionally, the winding number $\mathcal{W}$ is no longer limited to values $-1$, $0$, or $1$, and can be equal to $-2$ or $2$. 
This suggests an increase in the number of possible localized topological states in the ladder compared to the simple chain as expected.

Such behavior can be explained by analysing  the block matrix $\mathcal{A}(k)$ used to calculate the winding number $\mathcal{W}$ [see Eqs.~(\ref{eq.ham_off_diag}) and~(\ref{eq.winding})].
$\mathcal{W}$ is evaluated using the determinant of $\mathcal{A}(k)$, hence changing the basis does not affect the result.
By changing the order of the  tensor product basis $(\nu,\sigma)\rightarrow(\sigma,\nu)$, we can rewrite $\mathcal{A}(k)$ as:
\begin{eqnarray}
\mathcal{A}(k) &=& (\mathcal{E}_{k} \rho^{1} \sigma^{0} - \mu \rho^{0} \sigma^{0} - h \rho^{0} \sigma^{3} + \imath \mathcal{L}_{k} \rho^{1} \sigma^{2} \nonumber \\
&& - m_{0} \rho^{3} \sigma^{3} + \imath \Delta \rho^{0} \sigma^{2} ) \nu^0 - t_{\perp} \rho^{0} \sigma^{0}  \nu^{1} + \eta \rho^{0} \sigma^{1}  \nu^{2} \nonumber \\
&=& \mathcal{A}_0 \nu^0 - t_{\perp} \rho^{0} \sigma^{0}  \nu^{1} + \eta \rho^{0} \sigma^{1} \nu^{2},
\end{eqnarray}
where $\mathcal{A}_0$ is the off diagonal block of the 1D AFM chain~\cite{kobialka.sedlmayr.21}. 
Using the properties of block diagonal matrices we can simplify the determinant of the block diagonal matrix $\mathcal{A}$
\begin{eqnarray}
\det(\mathcal{A}) &=& \det(\mathcal{A}_0) \det \left( \mathcal{A}_0 - T \mathcal{A}_0^{-1} T^{\dagger} ,\right)
\end{eqnarray}
where $T = - t_{\perp} \rho^{0} \sigma^{0} - \imath \eta \rho^{0} \sigma^{1}$ and $T^{\dagger} = - t_{\perp} \rho^{0} \sigma^{0} + \imath \eta \rho^{0} \sigma^{1}$. Hence we get 
\begin{eqnarray}
\det(\mathcal{A}) &=& \det \left( \mathcal{A}_0^2 -  \mathcal{A}_0 T \mathcal{A}_0^{-1} T^{\dagger} \right).
\end{eqnarray}
Assuming  $\eta = 0$,  
\begin{eqnarray}
\nonumber \det(\mathcal{A}) &=& \det \left( \mathcal{A}_0^2 -  t_{\perp}^2 \rho^0 \sigma^0 \right) \\
\nonumber &=& \det \left( \mathcal{A}_0 - t_{\perp} \rho^0 \sigma^0 \right) \cdot \det \left( \mathcal{A}_0 + t_{\perp} \rho^0 \sigma^0 \right) \\
&=& \det [\mathcal{A}_0( \mu - t_{\perp})] \cdot \det [\mathcal{A}_0( \mu + t_{\perp})].
\label{eq.wind_comb}
\end{eqnarray}
This equality shows that in the absence of Rashba SOC between the wires ($\eta$) the phase diagram as well as the winding number can be reproduced by superimposing two single wire phase diagrams and their corresponding winding numbers. 
This immediately implies the existence of winding number $\mathcal{W}\pm 2$. 
The winding number is still an odd function of the chemical $\mu$, while an even function of the Zeeman field $h$, which was already reported for uncoupled 1D chains.

Features discussed above demonstrate the crucial role of the SOC $\eta$ in the topological phase diagram.
For example, the phase diagram can be tuned by the coupling between chains (via the hopping $t_{\perp}$ and the SOC $\eta$).
Nevertheless, the SOC $\eta$ leads to occurrence of new non-trivial parts in the phase diagram.
These properties can be observed by comparing  the phase diagrams with or without the SOC $\eta$.
An example is presented in Fig.~\ref{fig.hmu_ladder2} where
the main structure of the phase diagram remains unchanged (cf.  Fig.~\ref{fig.hmu_ladder}).
However, the introduction of the SOC $\eta$ leads to new topologically non-trivial regions around $\mu / t = \pm 2$ (cf.~insets on Fig.~\ref{fig.hmu_ladder2}).
Moreover, these regions can exist even for infinitesimal magnetic fields.
Similar behavior was already present in the case of the AFM chain close to the half-filling region (around $\mu = 0$)~\cite{kobialka.sedlmayr.21}.
However, contrary to the AFM chain,  higher winding numbers are found for ladders (like $\mathcal{W} = \pm 3$ in small topological region marked by red arrows on Fig.~\ref{fig.hmu_ladder2}).
Furthermore, we observed two effects in Fig.~\ref{fig.hmu_ladder3}:  i) a splitting of topological regions around half-filling and ii) the emergence of new topological regions close to the nearly empty/occupied band (i.e. close to the bottom/top of band, $\mu / t = \pm 2$).

Further tuning of the coupling between the chains can lead to  more complicated phase diagrams.
Indeed, the phase diagram for a larger interchain coupling ($t = t_{\perp}$ and $\lambda = \eta$) is presented in Fig.~\ref{fig.hmu_ladder3}.
Similar to the 1D chain, the non-trivial topological region exhibits parabolic boundaries for small $h$.
However, contrary to the simple chain, increasing magnetic field result in parts with larger winding numbers $\mathcal{W}\pm 2$ and $\mathcal{W}\pm 3$.
Additionally, close to $h / t = \pm 1$, we encounter a thin region of winding number $\pm 3$.
These regions exhibit complex substructures when examined closely [see Figs.~from~\ref{fig.hmu_ladder3}(a) to ~\ref{fig.hmu_ladder3}(d)].
In fact, we find in Fig.~\ref{fig.hmu_ladder3}(d) that the winding numbers  $\mathcal{W}=0$ to $\mathcal{W}=-3$ coexist in very close proximity.

The topological phase is very sensitive to the system parameters, and can exhibit fractal-like behavior.
In fact, similar behavior can be found for different sets of system parameters.
For example Fig.~\ref{fig.hmu_ladder4} show the phase diagram for $h / t = 0.975$, while magnetic moment value $m_{0}$ is parameter.
In this case, the nontrivial range of phase diagram with wild range of the winding number $\mathcal{W}$ is present.
We can notice that regions with larger $\mathcal{W}$ are realized in smaller range of parameters.
Nevertheless, for small values of magnetic moment $m_{0}$, the phase diagram exhibits fractal-like feature with comb-like structure [Fig.~\ref{fig.hmu_ladder4}(b)].
Parts of phases with $\mathcal{W} = -1$ and $-3$ are alternating along constant magnetization $m_{0} = \text{const.}$ cross-sections.

\begin{figure}[!t]
\centering
\includegraphics[width=\linewidth]{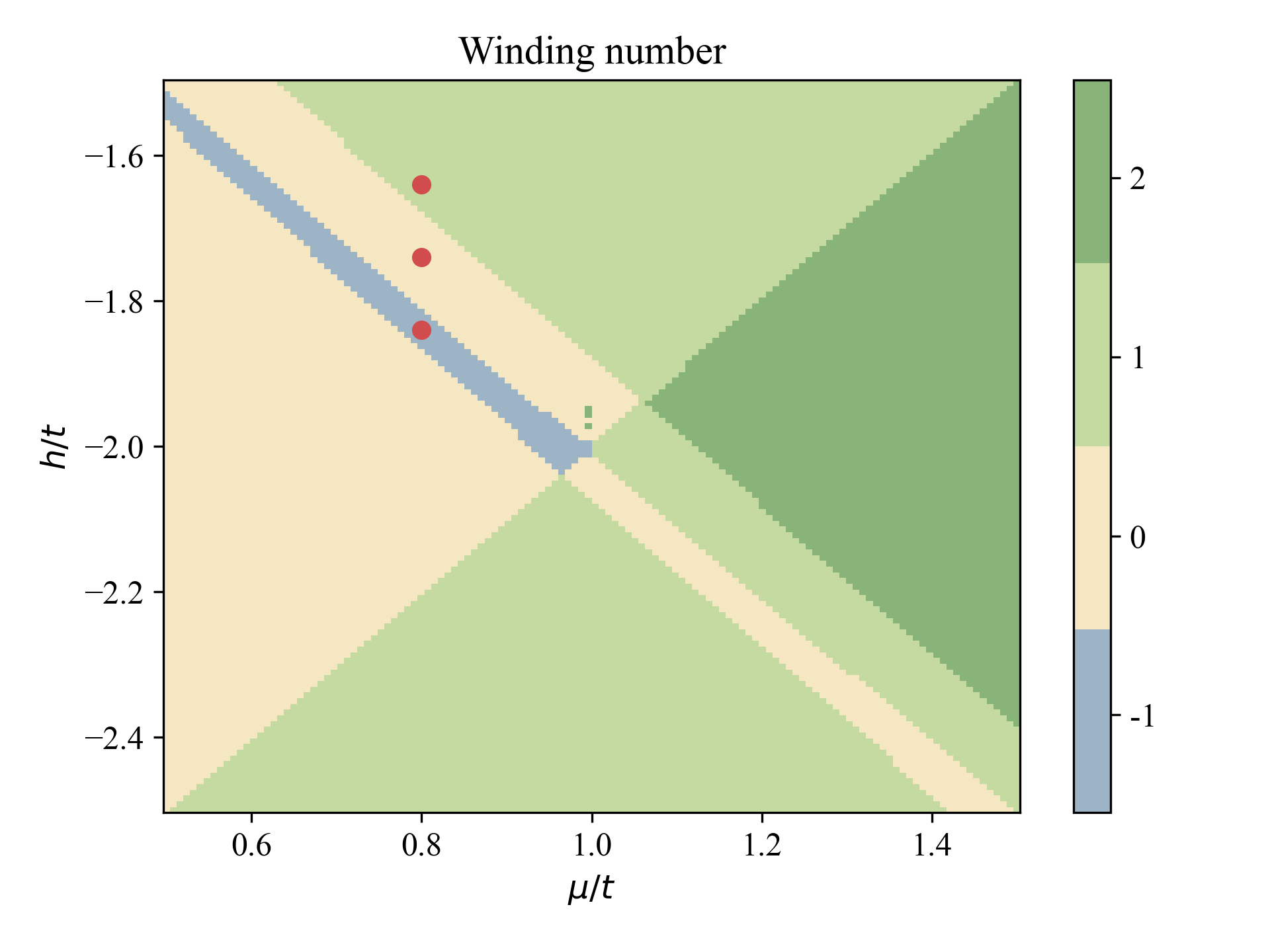}
\caption{The phase diagram of the winding number for $\mu$ vs. $h$. The magnetic moment $m_0 = -0.1 t$, superconducting order $\Delta = -0.2 t$, and as before $\lambda = \eta = -0.15 t$ along with $t = t_{\perp}$. The red dots are the points in the phase diagram which have been chosen to display the band inversion located at $\mu = 0.8$ and $h / t = -1.84$, $-1.74$, and $-1.64$ (bottom to top).
\label{fig.bandinv_phase}}
\end{figure}

\begin{figure*}
\centering
\includegraphics[width=\textwidth]{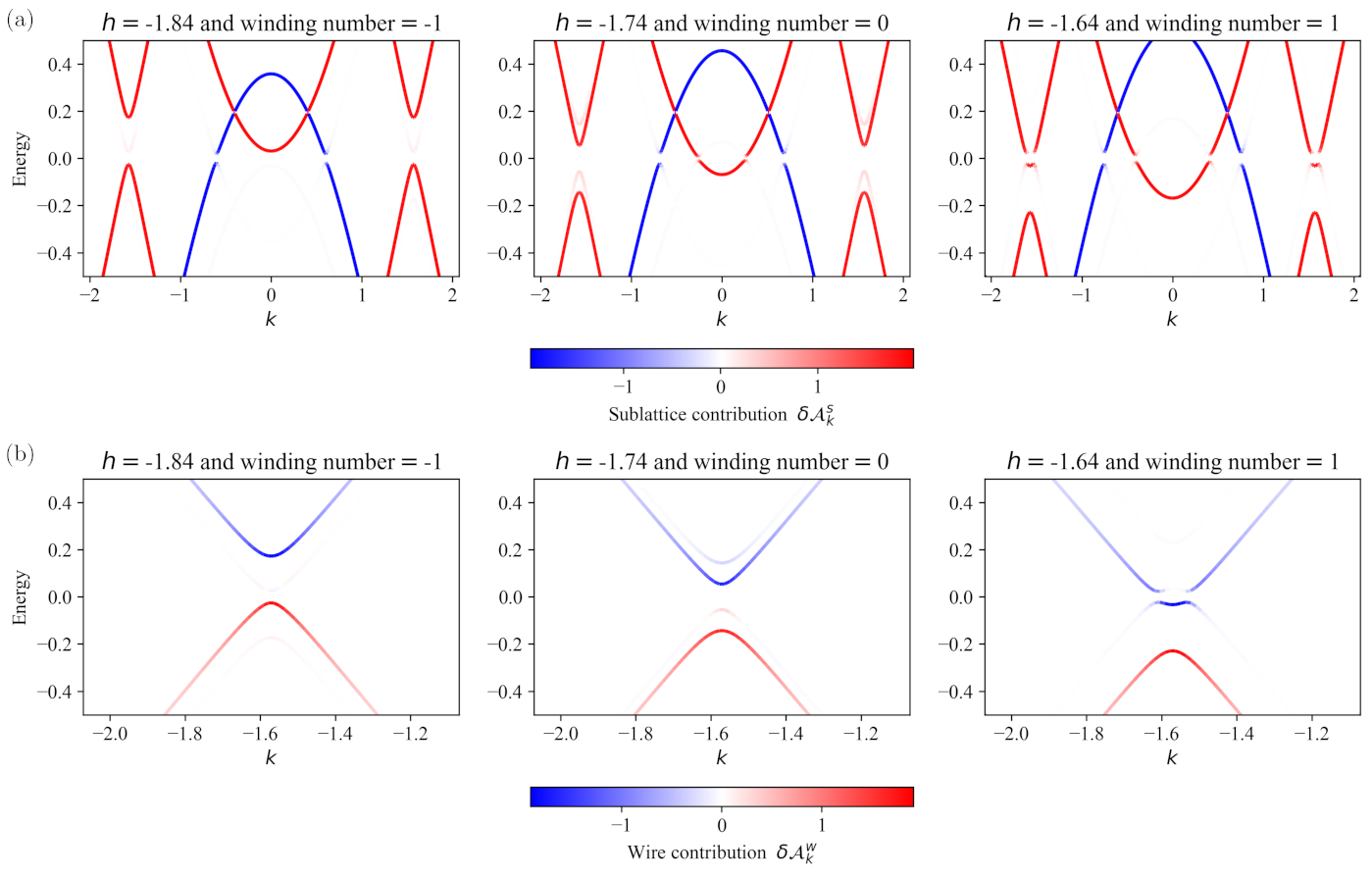}
\caption{ The energy bands graded with the differential spectral function for the (a) sublattice and (b) wire. The energy bands have been plotted for the parameter value marked red in Fig.~\ref{fig.bandinv_phase}
\label{fig.bandinv}}
\end{figure*}


\subsection{Band inversions and spectral function}

The transition from trivial to topological phase can be investigated within the band inversion formulation.
In fact, the system band structure is given by the eigenvalue problem of the Hamiltonian in the matrix form~(\ref{eq.ham_mom_matrix}).
Thus, bands can be investigated within different subspace (as mention in Sec.~\ref{sec.theo}) using the spectral function:
\begin{eqnarray}
\mathcal{S}_{\bm k} (\omega) &=& -\frac{1}{\pi} \sum_{s w \sigma} \text{Im} \; \mathcal{G}_{{\bm k} s w \sigma} \left( \omega + \imath 0^{+} \right),
\end{eqnarray}
where $\mathcal{G} = \left( \omega - \mathbb{H}_{\bm k} \right)^{-1}$ is the Green function.
The spectral function $\mathcal{S}_{\bm k} (\omega)$ can be then reexpressed in terms of the momentum space Bogoliubov--de Gennes (BdG) equations, as follows:
\begin{eqnarray}
&& \mathcal{S}_{\bm k} (\omega) = \sum_{s w \sigma} \mathcal{S}_{{\bm k} s w \sigma} (\omega) \\
\nonumber &=& \sum_{{\bm k} s w \sigma n} \left[ \vert u_{{\bm k} s w \sigma}^{n} \vert^2 \delta \left( \omega - \mathcal{E}_{n} \right) + \vert v_{{\bm k} s w \sigma}^{n} \vert^2 \delta \left( \omega + \mathcal{E}_{n} \right) \right] ,
\end{eqnarray}
where $u_{{\bm k} s w \sigma}^{n}$ and $v_{{\bm k} s w \sigma}^{n}$ are  particle and hole part of the $n$th eigenvector of the BdG equations of $\mathbb{H}_{\bm k}$~(\ref{eq.ham_mom_matrix}).
Here, $n$ denotes the band index  out of a total of 16 bands with energies $\mathcal{E}_{n}$.
The contribution of the different subspaces to different bands can be indicated by ($A$ and $B$) sublattice, ($\alpha$ and $\beta$) wire, ($\uparrow$ and $\downarrow$) spin-dependent (partial) spectral function:
\begin{eqnarray}
\delta \mathcal{S}_{\bm k}^{s} &=& \mathcal{S}_{{\bm k} A \alpha \uparrow} +  \mathcal{S}_{{\bm k} A \alpha \downarrow} + \mathcal{S}_{{\bm k} A \beta \uparrow} + \mathcal{S}_{{\bm k} A \beta \downarrow} \\
\nonumber && - \mathcal{S}_{{\bm k} B \alpha \uparrow} -  \mathcal{S}_{{\bm k} B \alpha \downarrow} - \mathcal{S}_{{\bm k} B \beta \uparrow} - \mathcal{S}_{{\bm k} B \beta \downarrow} , \\
\delta \mathcal{S}_{\bm k}^{w} &=& \mathcal{S}_{{\bm k} A \alpha \uparrow} +  \mathcal{A}_{{\bm k} A \alpha \downarrow} + \mathcal{S}_{{\bm k} B \alpha \uparrow} + \mathcal{S}_{{\bm k} B \alpha \downarrow} \\
\nonumber && - \mathcal{S}_{{\bm k} A \beta \uparrow} -  \mathcal{S}_{{\bm k} A \beta \downarrow} - \mathcal{S}_{{\bm k} B \beta \uparrow} - \mathcal{S}_{{\bm k} B \beta \downarrow} , \\
\delta \mathcal{S}_{\bm k}^{\sigma} &=& \mathcal{S}_{{\bm k} A \alpha \uparrow} +  \mathcal{S}_{{\bm k} B \alpha \uparrow} + \mathcal{S}_{{\bm k} A \beta \uparrow} + \mathcal{S}_{{\bm k} B \beta \uparrow} \\
\nonumber && - \mathcal{S}_{{\bm k} A \alpha \downarrow} -  \mathcal{S}_{{\bm k} B \alpha \downarrow} - \mathcal{S}_{{\bm k} A \beta \downarrow} - \mathcal{S}_{{\bm k} B \beta \downarrow} .
\end{eqnarray}
It is evident from the form of these differential spectral functions that each of these quantities measure the imbalance of the contribution from their respective subspace and therefore allows for a careful analysis and labelling of the different phases.
Additionally, the band polarization which refers to the absolute sign of the partial spectral function, $\delta \mathcal{S}_k$ changes as we cross boundary between different phases (in sense of the winding number).

\begin{figure*}
\centering
\includegraphics[width=\linewidth]{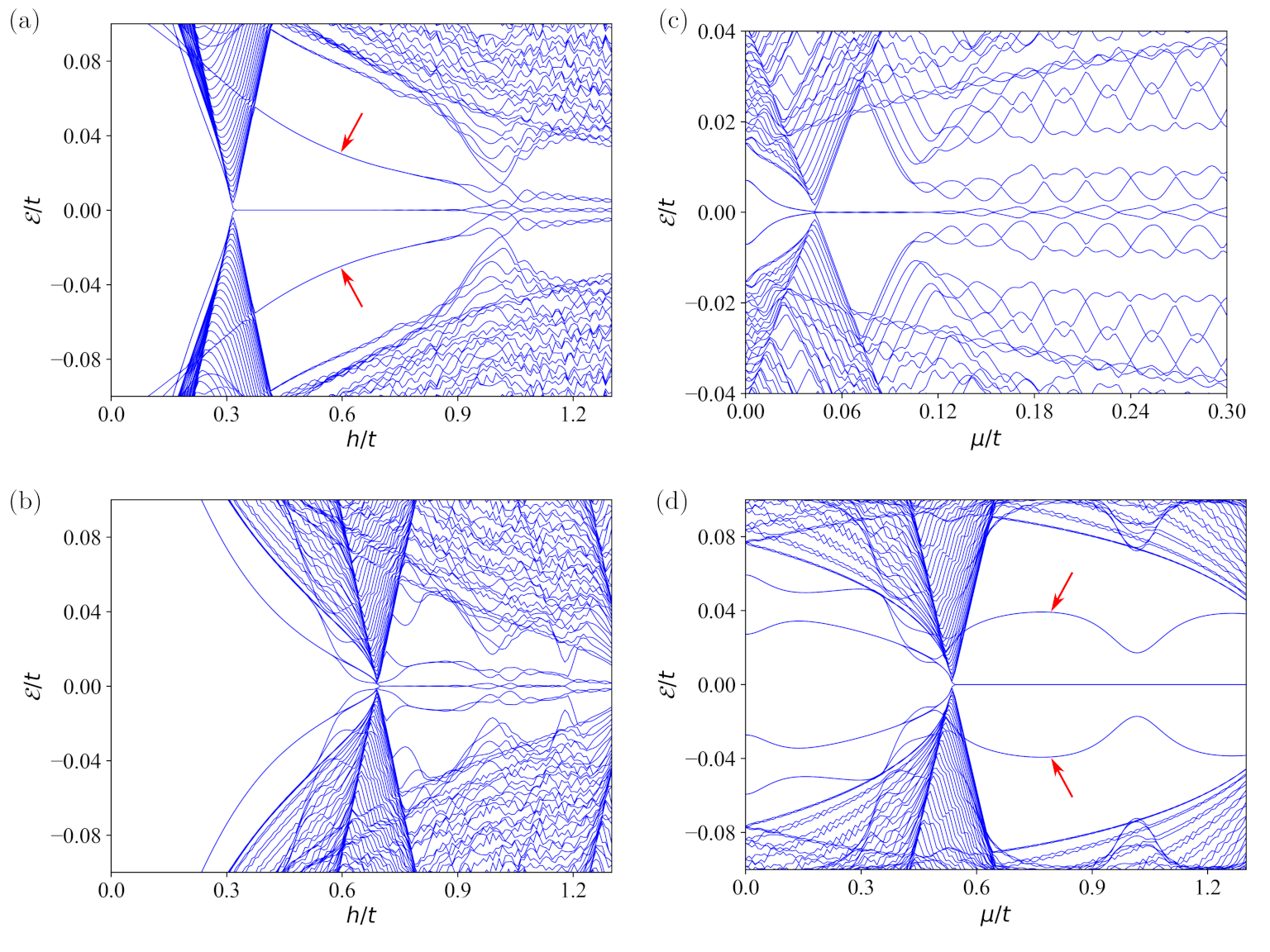}
\caption{Spectrum of the system with 201 sites along ladder, obtained from direct diagonalization of the Bogoliubove--de Gennes equations in real space. 
Results obtained for $m_0 / t = 0.1$, $\lambda = \eta = 0.15 t$, $\Delta / t = -0.2$, and $t_{\perp} = t$.
Results for fixed $\mu / t = 1/3$ (a), and fixed $h / t = 0.5$ (b).
\label{fig.band_disp}}
\end{figure*}

\begin{figure}[!t]
\centering
\includegraphics[width=\linewidth]{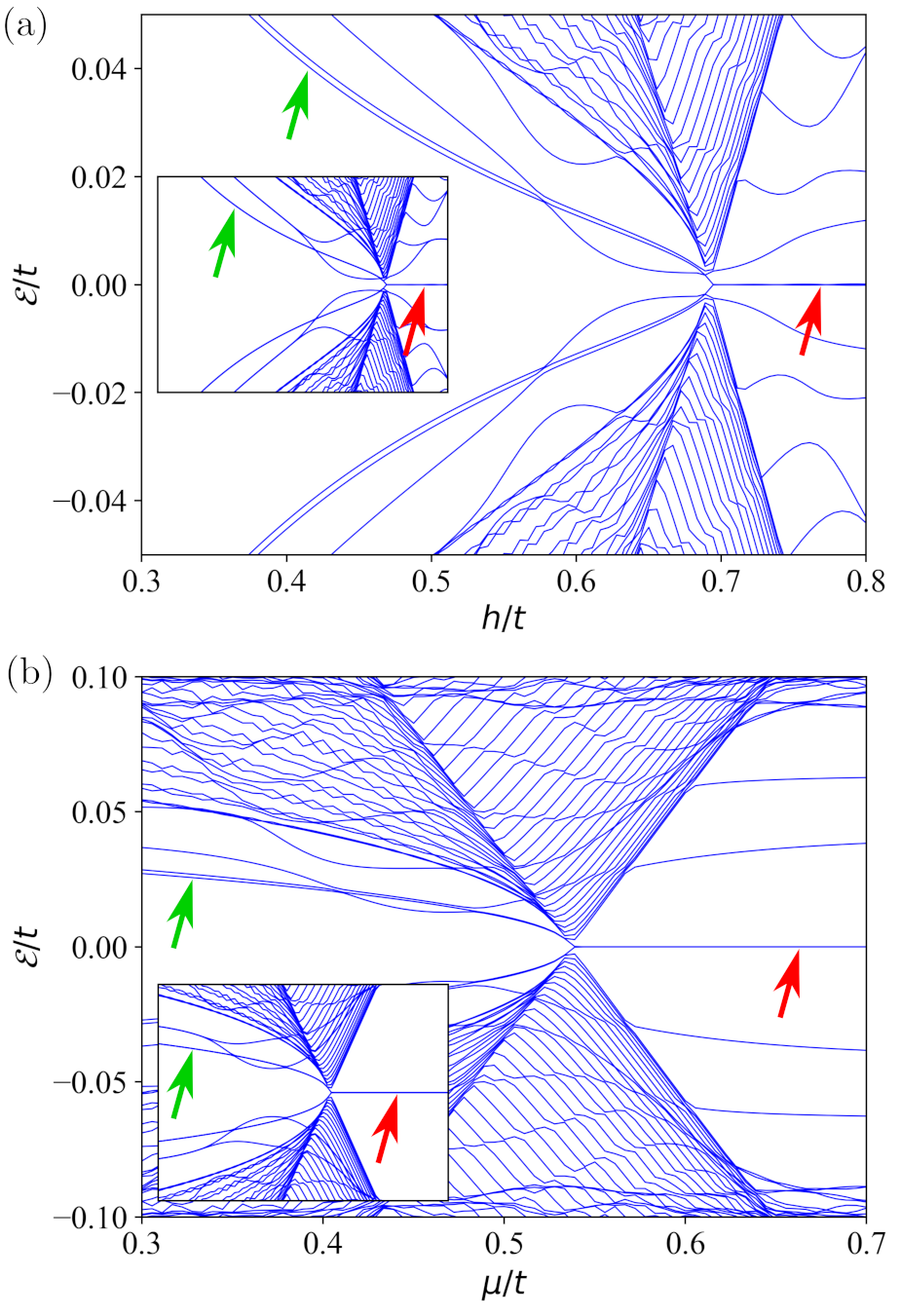}
\caption{The same as Fig.~\ref{fig.band_disp} for different lenght of the subchains.
Energy spectrum for 202 sites is presented on main panels, while insets show results for 201 sites.
Results obtained for $m_0 / t = 0.1$, $\lambda / t = \eta / t = 0.15$, $\Delta / t = -0.2$, and $t_{\perp} = t$.
Results for fixed $\mu / t = 1/3$ (a), and fixed $h / t = 0.5$ (b).
\label{fig.band_disp_ins}}
\end{figure}

To illustrate the use of these differential spectral functions, we analyse the part of the phase diagram where we observe transitions between different phases in  Fig.~\ref{fig.bandinv_phase}.
We choose specific points marked by red dots in  Fig.~\ref{fig.bandinv_phase}, characterized by different winding numbers ($-1$, $0$, and $1$, for fixed chemical potential $\mu / t = 0.8$ and magnetic fields $h / t = -1.84$, $-1.74$, and $-1.64$, respectively).
Increasing magnetic field leads  to two phase transitions: from nontrivial to trivial phase ($-1 \rightarrow 0$), and from trivial to nontrivial phase ($0 \rightarrow 1$).
We begin with the sublattice contribution $\delta \mathcal{S}_{\bm k}^{s}$ which is shown in Fig.~\ref{fig.bandinv}(a).
Beyond the Lifshitz transition at ${\bm k} = 0$ (initially unoccupied cross the Fermi level) induced by increasing the magnetic field, we observed leaking of the sublattice contribution between bands at ${\bm k} = \pm \pi / 2$.
Thus, the sublattice contribution to the bands is changed from ($0$,$1$,$0$,$1$) to ($1$,$0$,$1$,$0$) to ($1$,$1$,$0$,$0$) for bands from lowest to highest energies around the Fermi level.
However, independent of $\mathcal{W}$, bands exhibit strong sublattice polarization without clear inversion.
Situation is better suited for interpretation in the case of the wire contribution $\delta \mathcal{S}_{\bm k}^{w}$ around ${\bm k} = \pm \pi /2$, presented on Fig.~\ref{fig.bandinv}(b).
Initially two bands exhibit strong wire polarization, which after increasing the magnetic field leads to modification of the subchain contribution from ($0$,$1$,$0$,$-1$) to ($1$,$0$,$-1$,$0$) to ($1$,$-1$,$0$,$0$).
In general we see that the band structure analyses is more complicated than in single chain, due to the more complex band structure (which now contain 16 sub-bands). However by help of the  differential spectral functions, one can understand these transitions.


\section{Real space investigation}

In this section we will investigate the edge modes in real space ladder with finite number of sites.
To diagonalize the real space Hamiltonian  [from Eqs.~(\ref{eq.ham_real_space_general}) to~(\ref{eq.hsc_real})], we use the Bogoliubov--Valatin transformation~\cite{degennes.18}:
\begin{eqnarray}
c_{i s w \sigma} = \sum_{n} \left( u_{i s w \sigma}^{n} \gamma_{n} - \sigma v_{i s w \sigma}^{n\ast} \gamma_{n}^{\dagger} \right) ,
\end{eqnarray}
where $\gamma_{n}$ and $\gamma_{n}^{\dagger}$ are the fermionic operators.
This transformation leads to a real space BdG equations in the form~\cite{balatsky.vekhter.06,
ptok.kobialka.17,cichy.ptok.20,kobialka.sedlmayr.21}:
\begin{eqnarray}
\label{eq.bdg_real} \mathcal{E}_{n} \Psi_{a}^{n} = \sum_{b} \mathbb{H}_{ab} \Psi_{b}^{n} ,
\end{eqnarray}
where $\mathcal{E}_{n}$ and $\Psi_{a = i s w \sigma}^{n} = \left( u_{i s w \sigma}^{n} \; v_{i s w \sigma}^{n} \right)$ denotes $n$th eigenpair of real space Hamiltonian $\mathbb{H}_{ab}$ in matrix form.
Using the BdG equations solution, the local density of states (LDOS) $\rho_{i s w \sigma} = - 1/\pi \; \text{Im} \; \langle \langle c_{i sw \sigma} \vert c_{i sw \sigma}^{\dagger} \rangle \rangle$ can be express as:
\begin{eqnarray}
\nonumber \rho_{i s w} &=& \sum_{n \sigma} \left[ \vert u_{i s w \sigma}^n \vert^2 \delta ( \omega - \mathcal{E}_{n} ) + \vert v_{i s w \sigma}^n \vert^2 \delta ( \omega + \mathcal{E}_{n} ) \right] \\
\label{eq.ldos} &=& \sum_n \rho_{i s w}^{n} .
\end{eqnarray}
Experimentally measured tunneling amplitudes using a scanning tunneling microscope (STM) gives the information about the edge modes in the system. 
The LDOS provides a theoretical path of gaining similar insights for our system.

In the real space description, the number of sites play an important role in the obtained results (see also Sec.~\ref{sec.even_odd})~\cite{kobialka.sedlmayr.21}.
In this part we investigate system with $201$ sites along ladder.
This gives the ladder with the same sublattice at both ends of system (i.e.~sequence \mbox{$A$-$B$-$\cdots$-$B$-$A$}).
We exactly diagonalize the Hamiltonian which gives the energy spectra for the system. In order to gain more insight about the bound states we vary the magnetic field $h$, or the chemical potential $\mu$, as shown in Fig.~\ref{fig.band_disp}.
In both cases, $h = \text{const.}$ and $\mu = \text{const.}$, across trivial and non-trivial phases.
Indeed, for example for Figs.~\ref{fig.band_disp}(a) and~\ref{fig.band_disp}(b), a typical behavior is observed:
The increase of  the magnetic field leads to the closing of the trivial gap and reopening of a new topological gap. 
Therefore, for some range of $h$, zero-energy in-gap Majorana modes are observed.
However, as we mentioned earlier, realization of the topological phase strongly depends on the system parameters. 
For example, around $h / t = 1$ at Fig.~\ref{fig.band_disp}(a), we observe lifting of the zero-modes degeneracy, and destruction of the topological phase.
Contrary to the 1D chain, the number of in-gap states is not limited to one pair with zero-energy, i.e. Majorana edge modes.
In some situation, additional in-gap states can be observed in the topological phase [marked by red arrow on Figs.~\ref{fig.band_disp}(a) or~\ref{fig.band_disp}(d)].

\subsection{Even--odd behavior}
\label{sec.even_odd}

The system spectrum shows drastic difference depending on the even--odd length of the ladder.
This is well visible when we compare the spectrum for system with comparable length. 
We choose 200 and 201 sites, and  the results are presented in Fig.~\ref{fig.band_disp_ins}.
The topological in-gap zero energy Majorana edge modes (marked by red arrows) are independent of the length, which directly show their topological nature.
However, there are also states, whose existence or degeneracy strongly depends on the system length.
Such even--odd behavior is directly associated with the magnetic order and therefore affects the observed LDOS.
For an odd number of sites, both edges cannot be distinguished via the site-inversion symmetry with respect to the mirror symmetry (when magnetic order realized sequence \mbox{A-B-$\cdots$-B-A} or sequence \mbox{B-A-$\cdots$-A-B}).
Similar symmetry cannot be realized in the case of an even number of sites (ladder with sequence \mbox{A-B-$\cdots$-A-B}).
Additionally, this even--odd behavior affects the degenerate states in the whole range of parameters (e.g. states marked by green arrow on Fig.~\ref{fig.band_disp_ins}).

\begin{figure*}
\centering
\includegraphics[width=\linewidth]{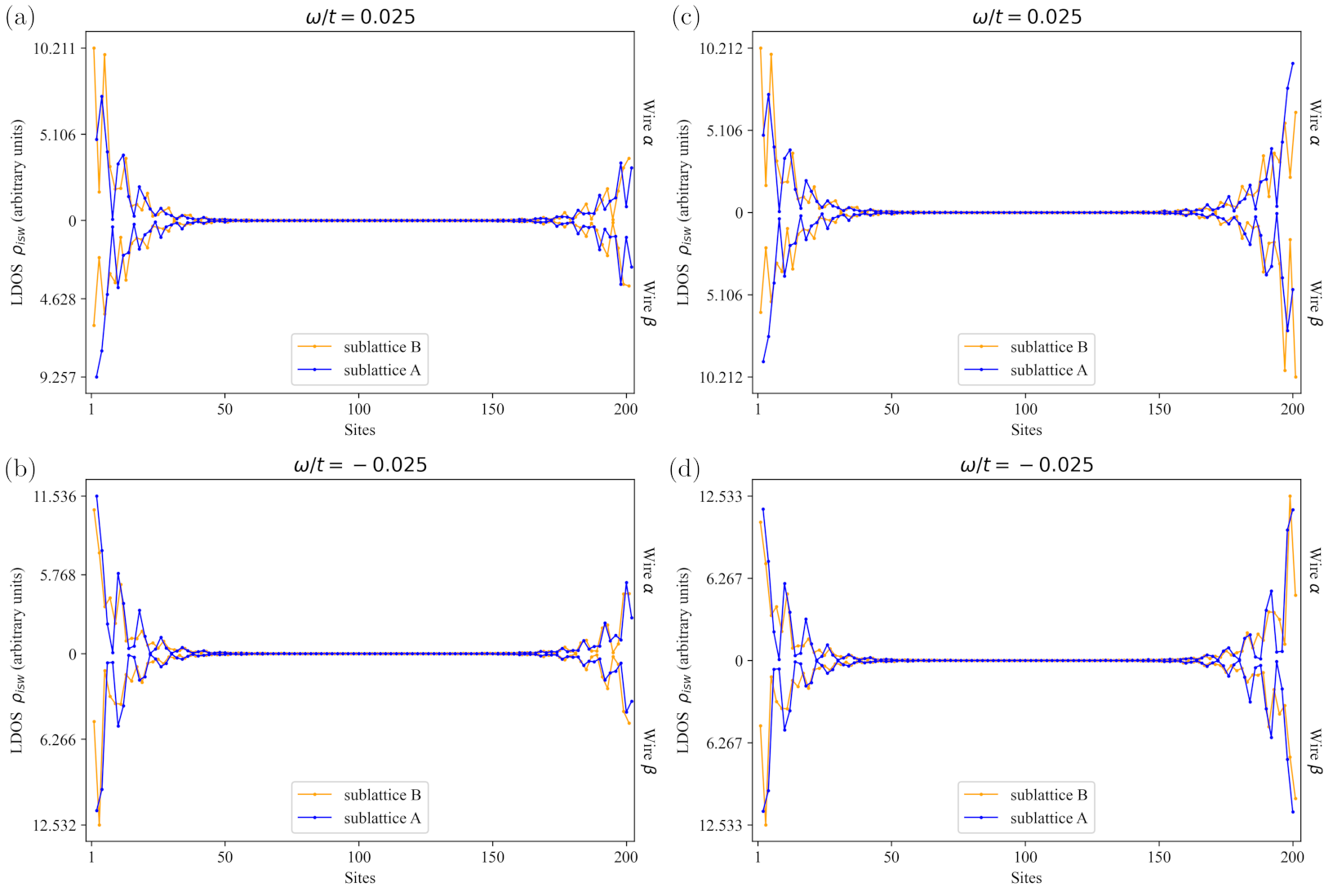}
\caption{ LDOS for even (left column) and odd (right column) number of sites on either wire. The LDOS is calculated for energy spectrum shown in Fig.~\ref{fig.band_disp_ins}(b) at $\mu = 0.35 t$. The LDOS is calculated in the trivial region of the spectra around or at the degenerated energy level i.e. $\omega = \pm 0.025 t$. 
\label{fig.ldos_triv1}}
\end{figure*}

\begin{figure}[!t]
\centering
\includegraphics[width=\linewidth]{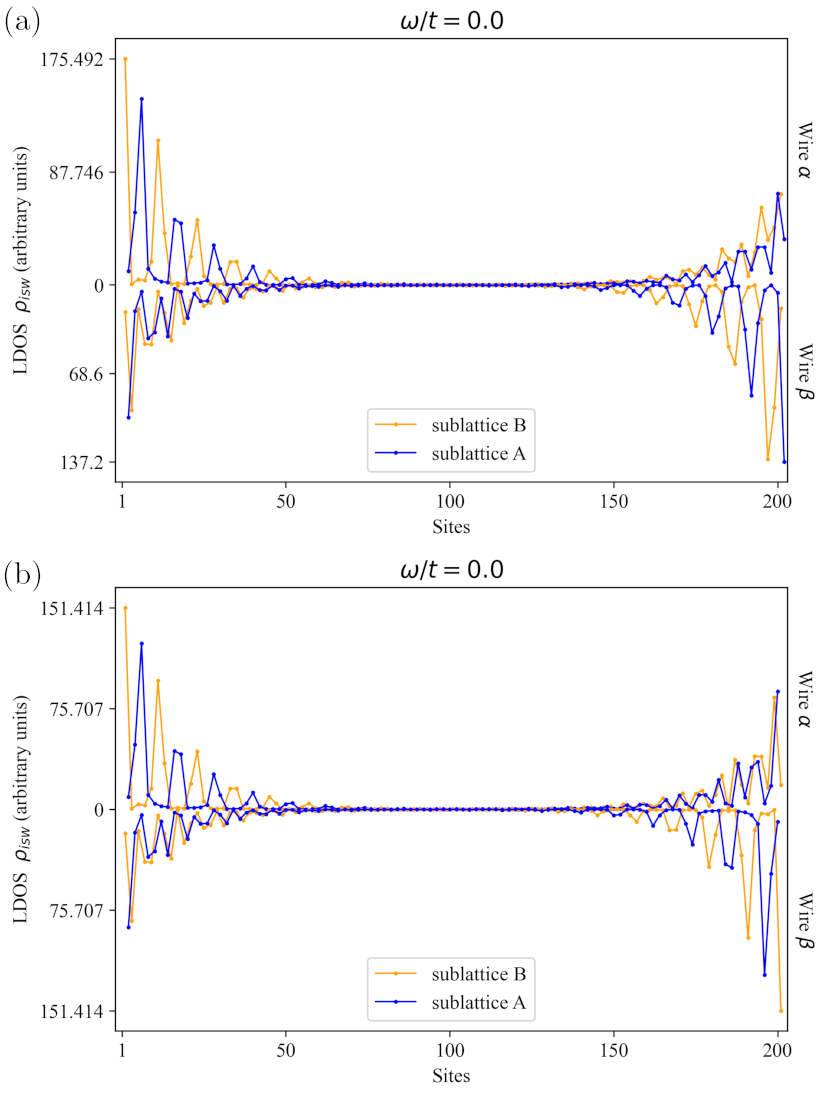}
\caption{ LDOS for topological bands at zero energy for the energy spectrum in Fig.~\ref{fig.band_disp_ins}(a) at $h = -0.8 t$ with even number of rungs in (a) and odd number of rungs in (b).
\label{fig.ldos_tp}}
\end{figure}

Similarly to the single AFM chain~\cite{kobialka.sedlmayr.21}, the magnetic field leads to the breaking of symmetry between left and right edge of the ladder resulting in the particle imbalance at both ends of the ladder.
The so-obtained LDOS are presented in Fig.~\ref{fig.ldos_triv1} for two pairs of double-degenerated non-zero in-gap states, while in Fig.~\ref{fig.ldos_tp} for one pair of in-gap zero energy Majorana edge modes. 
Top and bottom parts of panels, correspond to the subchain $\alpha$ and $\beta$, respectively, while blue and orange lines to sublattice A and B, respectively. 
As we mentioned earlier, for an even number of sites, there is strong difference between both sites [left panels on Fig.~\ref{fig.ldos_triv1}, and Fig.~\ref{fig.ldos_tp}(a)].
Contrary to this, a ladder with an odd number of sites exhibits left-right symmetry.
However, We should notice, that this symmetry is not preserved within the same lattice, but between subchains $\alpha$ and $\beta$, i.e. left (right) site of subchain $\alpha$ is the same as right (left) site of subchain $\beta$.
This is true for an odd number of sites, independent of the energies of discussed in-gap states [right panels in Fig.~\ref{fig.ldos_triv1}, and Fig.~\ref{fig.ldos_tp}(b)].

This observation can be associated with the existence of a center of crystalline symmetry (an inversion) in the case of an odd number of sites along the subchains.
However, this center of crystalline symmetry do not only affect sites number within the (sub)chains, like in the AFM chain~\cite{kobialka.sedlmayr.21}, but also the subchains itself, i.e. $\alpha \leftrightarrow \beta$ [see Sec.~\ref{sec.even_odd}].
Thus, the left end of the $\alpha$ ($\beta$) subchain, can be transformed to right end of $\beta$ ($\alpha$) subchain, with simultaneous changes of the SOCs directions ($\lambda \rightarrow -\lambda$ and $\eta \rightarrow - \eta$).
The magnetic moments in A and B sublattice, as well as direction magnetic field $h$ or spin $\uparrow$/$\downarrow$ quantization axis, are unchanged.

Contrary to the zero energy modes the degenerate massive modes are not topologically protected.
Albeit, in the limit where the number of rungs in the ladder tend to infinitum, the topological phase illustrates that there are parameters regimes with non-zero winding number.
The absence of corresponding additional zero-energy modes in the finite system suggests a transition from multiple Majorana modes (infinite ladder) to phases with a single Majorana mode and massive fermionic modes (finite ladder).

\section{Summary and outlook}
\label{sec.sum}

Topological phase realized in  chains with ferromagnetic or antiferromagnetic order in proximity of a superconductor can lead to  topologically protected Majorana edge modes.
Recently, such feature was discussed in several systems where topological phases were observed in wide ranges of system parameters. 

In this paper we have investigated the topological phase diagrams  of a block magnetic ladder.
In the case of weak coupling between subchains, the topological phase diagram is very similar to the one reported earlier for antiferromagnetic chains.
However, a topological phase exist in doubled region exhibiting negative or/and positive interference. 
Consequently, absolute value of the winding number is not limited to $\leq 1$ like in chains, and is $\leq 3$ in our model of the block ladder.
Additionally, the SOC between chains is responsible for the realization of new topological phases.
Similarly, like in the antiferromagnetic chain, the topological phase can exist even in the absence of the magnetic field, or around half-filling (far from bottom or top of band).

For strongly coupled  subchains, the topological phase diagrams becomes more complex.
and  is very sensitive to the system parameters. 
Additionally,  the topological phase diagram often exhibits a fractal-like structure around boundaries between phases with different winding numbers.
Thus, contrary to the ferromagnetic chain, the condition for realization of the topological phase cannot be given analytically using a closed analytic formula.

Finally, we check that the topological phases are associated with the realization of  in-gap edge modes. 
Some of them can be classified as  zero energy Majorana edge modes. 
However, for  phases with absolute value of the winding numbers larger than $1$, we have found  additional  non-zero energy in-gap states localized at the edges of the subchains.
Finally, such in-gap states exhibit an even--odd behavior with respect to the number of sites within the subchains which is related to the presence or absence of inversion symmetry in the crystal.

\begin{acknowledgments}
Some figures in this work were rendered using {\sc Vesta}~\cite{momma.izumi.11} software. 
S.Y. and A.P. are grateful to Laboratoire de Physique des Solides in Orsay (CNRS, University Paris Saclay) for hospitality during a part of the work on this project.
S.Y. acknowledges financial support provided by the Polish National Agency for Academic Exchange NAWA  under the Programme STER- Internationalisation of doctoral schools, Project no BPI/STE/2023/1/00027/U/00001
We kindly acknowledge the support by the National Science Centre (NCN, Poland)  under Project No.~2021/43/B/ST3/02166.
\end{acknowledgments}

\appendix

\section{Inversion symmetry for ``odd'' ladder}
\label{sec.inv}

In this appendix we discuss the inversion symmetry in the system with odd number of sites along chains.
The inversion symmetry with respect to the center of symmetry, would flip both site and subchain.
This the site lcoated at site $i$ in subchain $\alpha$ is migrated to location $j = N + 1 - i$ in subchain $\beta$ (here $N$ is the total number of sites in each subchains $\alpha$ and $\beta$).
It is now fruitful to demonstrate the action of inversion operator acting on the Hamiltonian.
The BdG Hamiltonian in real space (see Eq.~\ref{eq.bdg_real}), can be represent as a block-diagonal matrix:
\begin{eqnarray}
\mathbb{H} = \left( \begin{array}{cc}
H & {\bm \Delta} \\ 
{\bm \Delta}^{T} & -H^{\ast}
\end{array} \right) ,
\end{eqnarray}
where $H$ and ${\bm \Delta}$ are block tri-diagonal matrices, and ${\bm \Delta}^{T} = - {\bm \Delta}$.
Using the same representation as described in previous sections, we have:
\begin{eqnarray}
H = \left( 
\begin{array}{ccccccc}
D_{1} & K & • & • & • & • & • \\ 
K^{\ast} & D_{2} & K & • & • & 0 & • \\ 
• & K^{\ast} & D_{3} & • & • & • & • \\ 
• & • & • & \ddots & • & • & • \\ 
• & • & • & • & D_{N-2} & K & • \\ 
• & 0 & • & • & K^{\ast} & D_{N-1} & K \\ 
• & • & • & • & • & K^{\ast} & D_{N}
\end{array} \right) ,
\end{eqnarray}
where $K = -t \nu^{0} \sigma^{0} - t_{\perp} \nu^{1} \sigma^{0} - \imath \lambda \nu^{0} \sigma^{1} - \eta \nu^{2} \sigma^{1}$ and $D_{i} = \mu \nu^{0} \sigma^{0} - h \nu^{0} \sigma^{3} + (-1)^{i} m_0 \nu^{0} \sigma^{3}$ subblock are $4 \times 4$ matrices (index $i$ denote site along ladder).
The blocks represent two sites one on each wire within the unit cell (hence half the unit cell).
We note that $K$ incorporates the hoppings and SOCs along and between subchains, while the sublattice information is built into the diagonal block $D_i$.
Additionally, superconductivity is described by ${\bm \Delta} = \imath \Delta \nu^0 \sigma^2 \mathbb{I}_{N}$.

Now, we are focused on the ladder composed of odd number of sites along subchains, what correspond to the magnetic order A-B-$\cdots$-B-A (or equivalently B-A-$\cdots$-A-B).
The finite system reals space Hamiltonian remains invariant under geometrical inversion applied along $\hat{x}$ and $\hat{y}$ directions (we assume subchains along $\hat{x}$), while the magnetic order is unchanged.
We observe that inversion $H \rightarrow H^{\ast}$, on the other hand leaving ${\bm \Delta}$ block unaffected.
Thus, we can find the following operation:
\begin{eqnarray}
\mathcal{I} \mathbb{H} \mathcal{I}^{\dagger} = \left( \begin{array}{cc}
H^{\ast} & {\bm \Delta} \\ 
{\bm \Delta}^{T} & -H
\end{array} \right) ,
\end{eqnarray}
where $\mathcal{I}$ is the inversion operation given by:
\begin{eqnarray}
\mathcal{I} &=& 
\tau^{0} \times 
\left( \begin{array}{ccccc}
• & 0 & • & \sigma^0 \nu^{1} & • \\ 
• & • & \reflectbox{$\ddots$} & • & • \\ 
• & \sigma^0 \nu^{1} & • & 0 & • 
\end{array} \right) .
\end{eqnarray}
Consequently, we can find the Hamiltonian of the ladder after inversion $\mathbb{H}^\text{inv}$, as:
\begin{eqnarray}
\nonumber \mathbb{H}^\text{inv} = \mathcal{I} \mathbb{H} \mathcal{I}^{\dagger} = 
- \left( \begin{array}{cc}
- H^{\ast} & - {\bm \Delta} \\ 
- {\bm \Delta}^{T} & H
\end{array} \right) 
= - \left( \begin{array}{cc}
- H^{\ast} & {\bm \Delta}^{T} \\ 
{\bm \Delta} & H
\end{array} \right) \\
\end{eqnarray}
Such relation, combined with build-in particle-hole symmetry, implies that the inversion operation does not changes the spectrum.

\vspace{1cm}

\bibliography{biblio.bib}

\end{document}